\documentstyle[prl,aps]{revtex}
\input{psfig.sty}
\draft
\begin{document}
\twocolumn[\hsize\textwidth\columnwidth\hsize\csname@twocolumnfalse\endcsname

\title{Continuum  description of avalanches in granular media}
\author{Igor S. Aranson$^1$  and Lev S. Tsimring$^2$}
\address{
$^1$ Argonne National Laboratory,
9700 South  Cass Avenue, Argonne, IL 60439 \\
$^2$ Institute for Nonlinear Science, University of California,  
San Diego, La Jolla, CA 92093-0402 }
\date{\today}

\maketitle

\begin{abstract} 
We develop a continuum description of partially fluidized granular
flows.  Our theory  is based on the hydrodynamic equation for the
flow coupled with the order parameter equation which describes the
transition between flowing and static components of the granular
system.  This theory captures important phenomenology recently
observed in experiments with granular flows on rough inclined planes
(Daerr and Douady, Nature (London) {\bf 399}, 241 (1999)): layer
bistability, and transition from triangular avalanches propagating
downhill at small inclination angles to balloon-shaped avalanches
also propagating uphill for larger angles.  
\end{abstract}


\narrowtext
\vskip1pc]

Fundamental understanding of the dynamics of granular media still poses
a challenge for physicists \cite{jnb,kadanoff,gennes} and engineers
\cite{nedderman}.  The intrinsic dissipative nature of the interactions
between the constituent macroscopic particles sets granular matter apart
from conventional gases, liquids, or solids. One of the most interesting
phenomena pertinent to the granular systems is the  transition from a
static equilibrium to a granular flow. The most spectacular
manifestation of such a transition occurs during an avalanche. There has
been a number of experimental studies of avalanche flows in large
sandpiles\cite{bagnold,radj} as well as in thin layers of grains on rough
inclined surfaces \cite{daerr,daerr1,pouliquen}.  

On the theoretical side, a significant progress had been achieved by
large-scale molecular dynamics simulations \cite{ertas,walton} and by
continuum theory \cite{gennes1,bcre,gennes2,boutreux}.  The current
continuum approach  to the description of avalanche flows in the physics
community was pioneered by Bouchaud, Cates, Ravi Prakash and Edwards
(BCRE)\cite{bcre} and subsequently developed by de Gennes, Boutreux and
and Rapha\"el \cite {gennes1,gennes2,boutreux}.  In their model is the
granular system is spatially separated into two phases, static and
rolling.  The interaction between the phases is implemented through
certain conversion rates.  This model described certain features of
thin near-surface granular flows including avalanches. However, due to
its intrinsic assumptions, it only works when the granular material is well
separated in a thin surface flow and an immobile bulk. In many
practically important situations, this  distinction between ``liquid''
and ``solid'' phases is more subtle and itself is controlled by the
dynamics. 

In this Letter we propose a new continuum model for multi-phase granular
matter. The underlying idea of our approach is borrowed from the Landau
theory  of phase transitions\cite{landau}. We assume that the shear
stresses in a partially fluidized granular matter are composed of two
parts: the dynamic part proportional to the shear strain, and the
strain-independent (or ``static'')  part. The relative magnitude of the
static shear stress is controlled by the order parameter (OP) which
varies from 0 in the ``liquid'' phase to 1 in the ``solid'' phase.
Unlike ordinary matter, the phase transition in granular matter is
controlled not by the temperature, but the dynamics stresses themselves.
In particular, the Mohr-Coloumb yield failure condition\cite{nedderman} is
equivalent to a critical melting temperature of a solid.  The OP can be
related to the local entropy \cite{edwards} of the granular material.
OP dynamics is then coupled to the hydrodynamic equation for the
granular flow. We apply this model to study the  transition to flow in
thin granular layer on inclined planes with rough bottom. Our model
captures important phenomenology observed  by Pouliquen\cite{pouliquen}
and  Daerr and Douady\cite{daerr}, including the structure of the
stability diagram, triangular shape of downhill avalanches at small
inclination angles and balloon shape of uphill avalanches for larger
angles.  

{\it Model}. 
The continuum description of the granular flow is based on the 
Navier-Stokes equation
\begin{equation}
\rho_0 D v_i/Dt=\frac{\partial
\sigma_{ij} }{\partial x_j}+ \rho_0  g_i, \;\;j=1,2,3.
\label{elastic}
\end{equation}
where $v_i$ are the components of velocity, $\rho_0=const$ is the
density of material (we set $\rho_0=1$), ${\bf g}$ is acceleration of
gravity, and $D/Dt=\partial_t+v_i\partial_{x_i}$ denotes material
derivative. Since the relative density fluctuations are small, the
velocity obeys the incompressibility condition $\nabla\cdot {\bf v}=0$. 

The central conjecture  of our theory is that in partially fluidized
flows, some of the grains are involved in plastic motion, while others
maintain prolonged static contacts with their neighbors. Accordingly, we
write the stress tensor as a sum of the hydrodynamic part proportional
to the flow strain rate $e_{ij}$, and the strain-independent  part,
$\sigma_{ij}^s$, i.e.  $\sigma_{ij}=e_{ij}+\sigma_{ij}^s$.  We assume
that the diagonal elements of the tensor $\sigma_{ii}^s$ coincide with
the corresponding components of the ``true''  static stress tensor
$\sigma_{ii}^0$ for the immobile grain configuration in the same
geometry, and the shear stresses  are reduced by the value of the order
parameter  $\rho$ characterizing the ``phase state'' of granular
matter. Thus, we write the stress tensor in the form
\begin{equation}
\sigma_{ij}=
\eta \left(\frac{\partial v_i }{\partial x_j}  +
\frac{\partial v_j }{\partial x_i} \right) +
\sigma_{ij}^0\left(\rho+(1-\rho)\delta_{ij}\right).
\label{sigma}
\end{equation}
Here $\eta$ is the viscosity coefficient.  In a  static state,
$\rho=1$, $\sigma_{ij}=\sigma_{ij}^0$, $v_i=0$, whereas in a fully
fluidized state $\rho=0$, and the shear stresses are simply proportional
to the strain rates as in ordinary fluids.

To close the system we need a set of constitutive relations between
static shear and normal stresses, as well as an equation for the order
parameter $\rho$. The issue of constitutive relations in granular
materials is complex and not completely
understood\cite{nedderman,goddard}. It appears that in many cases, the
constitutive relations are determined by the construction
history\cite{wittmer}.  Recent studies indicated a fundamental
role of the network of the force chains which carry forces
longitudinally\cite{bouch1}.  We will assume that for any given
problem, the corresponding static constitutive relations have been
specified.

For the order parameter $\rho$, we apply pure dissipative  dynamics
which can be derived from the ``free-energy'' type functional ${\cal
F}$, i.e., $\dot{\rho} = -\delta {\cal F}/\delta \rho$.  We adopt the
standard Landau form for ${\cal F} \sim \int {\bf dr }  ( D | \nabla
\rho|^2 + f(\rho,\phi))$, which includes a ``local potential energy''
and the diffusive spatial coupling.  The potential energy $f(\rho,\phi)$
should have extrema at $\rho=0$ and $\rho=1$ corresponding to uniform
solid and liquid phases.  According to the Mohr-Coulomb yield criterion
for non-cohesive grains\cite{nedderman} or its generalization
\cite{bouch1}, the static equilibrium failure and transition to flow is
controlled by the value of the non-dimensional ratio
$\phi=\max|\sigma_{mn}^0/\sigma_{nn}^0|$, where the maximum is sought over
all possible orthogonal directions $n$ and $m$ in the bulk of the granular
material. We simply use this ratio as a parameter in the potential energy
for the OP $\rho$.  Without loss of generality, we write the equation
for $\rho$:
\begin{equation}
\dot{\rho} =D\nabla^2\rho-a\rho(1-\rho)F(\rho,\phi)
\label{op-eq}
\end{equation}
Further, according to observations we assume that the static equilibrium
is unstable if $\phi\le \phi_1$, where $\tan^{-1}\phi_1$ is the
internal friction angle for a particular granular material.
Additionally, we assume that if $\phi<\phi_0$, the ``dynamic'' phase
$\rho=0$, is unstable.  Values of $\phi_0$ and $\phi_1$ do not coincide
in general. Typically there is a range in which both static and dynamics
phases co-exist (this is related to the so-called Bagnold
hysteresis\cite{bagnold}).  The simplest form of $F(\rho,\phi)$ which
satisfies these constraints, is $F(\rho,\phi)=-\rho+\delta$, where
$\delta=(\phi-\phi_0)/(\phi_1-\phi_0)$.  Setting  $D=1$ and $a=1$ we
arrive at 
\begin{equation}
\dot{\rho}
=\nabla^2\rho+\rho(1-\rho)(\rho-\delta).
\label{op-eq1}
\end{equation}
For  $\phi_0<\phi<\phi_1$ both static ($\rho=1$) and dynamic ($\rho=0$) phases are
linearly stable, and Eq.(\ref{op-eq1}) possesses a moving front solution
which ``connects'' these phases. The speed of the front in the direction
of $\rho=0$ is given by $V=(1-2\delta)/\sqrt{2}$.  At $\delta=1/2$ 
both phases co-exist.

\begin{figure}[h]
\centerline{ \psfig{figure=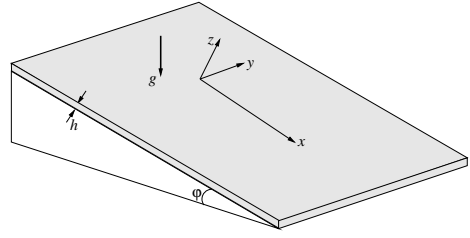,height=1.2in}}
\caption{Schematic representation of a chute geometry}
\label{setup}
\end{figure}

{\em Chute flow.} Let us now apply this 
formulation to a specific problem
of the chute flow. We consider a layer of dry cohesionless grains
on an inclined rough surface (see Fig.\ref{setup}). 
In the static equilibrium one has the following conditions: 
\begin{eqnarray}
\label{eqv}
\sigma_{zz,z}^0+ \sigma_{xz,x}^0 = -g \cos \varphi \;,\;
\sigma_{xz,z}^0+ \sigma_{xx,x}^0 = g \sin \varphi
\end{eqnarray}
where the subscripts after commas mean partial derivatives.  The
solution to Eqs. (\ref{eqv}) in the absence of lateral stresses
$\sigma^0_{yy}= \sigma^0_{yx}=\sigma^0_{yz}=0, $ is given by 
\begin{eqnarray} 
\label{eqv1} 
\sigma_{zz}^0=-g \cos \varphi\,z  \;,\;
\sigma_{xz}^0=g \sin \varphi\,z\;,\;\sigma_{xx,x}^0=0 
\end{eqnarray} 

In a static equilibrium there is a simple relation between shear and
normal stresses, $\sigma_{xz}^0 = -  \tan \varphi \sigma_{zz}^0$.
According to our conjecture, this relation between the static components
of the stress is maintained in the flowing regime as well.  For the
chute flow geometry, the value of parameter $\phi$ in Eq. (\ref{op-eq1})
can also be easily specified.  In this case, the most ``unstable'' yield
direction is parallel to the inclined plane, so  we can simply write
$\phi=|\sigma_{xz}^0/\sigma_{zz}^0|$. 

{\it Stationary solutions} of  Eq. (\ref{op-eq1}) for the confined chute
geometry Fig. \ref{setup} are subject to the following boundary
conditions (BC):  no-flux condition $\rho_z= 0$ at the free surface $z=0$,
and  $\rho=1$ at the bottom of the chute $z=-h$ (a granular medium is
assumed to be in a solid phase near the rough surface).  
There always exists a stationary solution to Eq. (\ref{op-eq1})
$\rho=1$ corresponding to a static equilibrium.  For $\delta>1$
it is stable at small
$h$, but loses stability at a certain threshold $h_c>1$. The most
``dangerous" mode of instability satisfying the above boundary
conditions, is $a\cos(\pi z/2h)$. The eigenvalue of this mode is
$\lambda(h)=\delta-1-\pi^2/4h^2$, hence the neutral curve $\lambda=0$
for the linear stability of the solution  $\rho=1$ is given by
\begin{equation}
h_c=\frac{\pi}{2\sqrt{\delta-1}}.
\label{stab1}
\end{equation}
For $h>h_c(\delta)$ grains spontaneously start to roll, and a granular
flow ensues.  In addition to the trivial state $\rho=1$, for 
$h>h_s(\delta)$ there exists a unique non-trivial stationary
solution satisfying the above BC. The value of $h_s$
can be found as a minimum of the following integral as a function of
$\rho_0$, the value of $\rho$ at the surface $z=0$,
\begin{equation}
h_s=\min \int_{\rho_0}^1 \frac{d\rho}{\sqrt{\frac{\rho^4}{2}-
\frac{2 (\delta+1)\rho^3}{3} +\delta\rho^2-c(\rho_0)}},
\label{hmin}
\end{equation}
where $c(\rho_0)=\rho_0^4/2-2 (\delta+1)\rho_0^3/3+\delta\rho_0^2$.
This integral can be calculated analytically for 
$\delta \to \infty$ and $\delta\to 1/2$. 
It is easy to show that for large $\delta$, 
the critical solution of Eq.(\ref{op-eq1}) has a form
$\rho= 1+a\cos(k z)$ with  $a\ll 1$ and $k=(\delta-1)^{1/2}$, and 
therefore, $h_s(\delta)\to h_c(\delta)$.  For
$\delta \to 1/2$, the critical phase trajectory comes close to two
saddle points $\rho=0$ and $\rho=1$, and an asymptotic evaluation of
(\ref{hmin}) gives $h_s = -\sqrt 2 \log (\delta-1/2) + const$.  This
expression agrees with the empirical formula $\phi - \phi_0  \sim
\exp[-h_s/h_0]$ proposed in Ref. \cite{daerr}.

\begin{figure}[h]
\centerline{ \psfig{figure=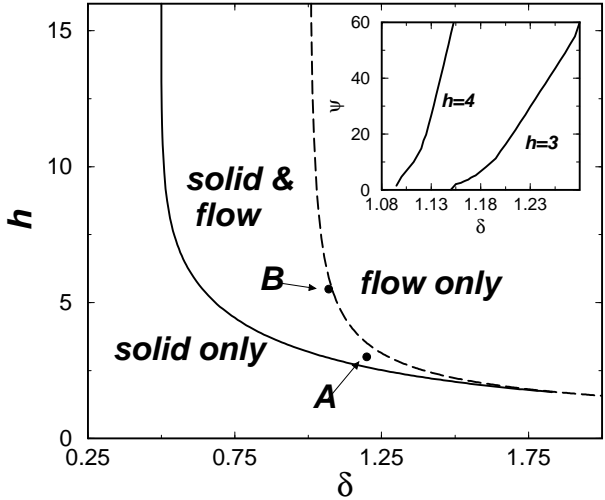,height=3.in}}
\caption{Stability diagram. 
Dashed line shows  the  neutral curve (\protect\ref{stab1}),
solid line shows the existence limit of fluidized state  (\protect\ref{hmin}).
Inset:  $\psi$ (in degrees) vs $\delta$ for $ \alpha=0.15 $ 
and $\beta=0.25$}
\label{stab}
\end{figure}

Neutral stability curve $h_c(\delta)$ and the critical line $h_s(\delta)$
limiting the region of existence of non-trivial granular flow solutions,
are shown in Fig.\ref{stab}.  They divide the parameter plane
$(\delta,h)$ in three regions. At $h<h_s(\delta)$, the trivial static
equilibrium $\rho=1$ is the only stationary solution of
Eq.(\ref{op-eq1}) for chosen BC.  For
$h_s(\delta)<h<h_c(\delta)$, there is a bistable regime, the static
equilibrium state co-exists with the stationary flow.  For
$h>h_c(\delta)$, the static regime is linearly unstable, and the only
stable regime corresponds to the granular flow. This qualitative picture
completely agrees with the recent experimental
findings\cite{daerr,pouliquen}.  Moreover, for rough bottom BC
(corresponding to our $\rho=1$), authors of Ref.\cite{daerr}
found a region of bistability in the parameter plane $(h,\varphi)$
which has a shape very similar to our phase diagram Fig.\ref{stab}.

The velocity profile corresponding to a stationary profile of $\rho(z)$,
can be easily found from Eq. (\ref{sigma}),
\begin{equation} 
\eta \frac{\partial v_x}{\partial z} =  g \sin \varphi z - \rho\sigma
_{xz}^0= g \sin \varphi (1-\rho)z.
\label{vel}
\end{equation} 
The  flux of grains  in the stationary flow $J$ is given by
\begin{eqnarray}
J= \int _{-h}^0 v_y(z) dz = 
\frac{g \sin \varphi }{\eta} \int _{-h}^0 \int _{-h}^z
(1-\rho(z^\prime)) z^\prime d z^\prime d z 
\label{j}
\end{eqnarray}

Pouliquen \cite{pouliquen} proposed a scaling for the mean velocity
$\bar v=J/h$  vs  thickness
of the layer $h$ in the stationary flow regime, $\bar v\propto
h^{3/2}/h_s$, which works for  angles $\varphi$ as well as
for different grain sizes.  Eq. (\ref{j})  yields  $v\propto
(h-h_s)^{1/2}$ for small $h-h_s$ and $v\propto h^2$ for large $h$. It
is plausible 
that the experimentally found scaling exponent $3/2$ is the
result of the crossover between two different regimes. However,
renormalization $\bar v/\sqrt{gh}, h/h_s$ as in Ref.\cite{pouliquen}
does not collapse our results onto a single curve, perhaps 
due to the assumption of a simple Newtonian
relation between the strain $v_z$ and the hydrodynamic part of the shear 
stress $\sigma_{xz}$ with a fixed viscosity  $\eta$ (see Eq.(\ref{sigma})). 
In fact,  $\eta$ itself may depends on $\rho$ and $z$ in some fashion. 


%


For a deep chute ($h\gg1$),  the  stationary solution  of
Eq.(\ref{op-eq1}) can be found analytically (cf. Ref.\cite{akv}).
However, in this case  the slope of the free surface may not be equal to
the slope of the inclined plane, but is itself determined by the amount of sand
which is poured on the surface up-stream.  
Thus,  the closure of the problem will be  provided by the constraint
$J=const$. 

{\it Avalanches in shallow chute}. 
In the vicinity of the neutral curve (\ref{stab1})
Eqs.(\ref{elastic},\ref{op-eq}) can be simplified.  We look for solution
in the form 
\begin{equation} 
\rho =1 - A \cos\left(\frac{\pi}{2 h} z\right)+ \mbox{h.o.t.},
\label{form1} 
\end{equation} 
where $A\ll 1$ is a slowly varying function of $t,\ x$, and $y$. 
Substituting ansatz  (\ref{form1}) into Eq. (\ref{op-eq}) and applying
orthogonality conditions, we obtain 
\begin{equation} 
A_t = \lambda(h) A+ \nabla^2_\perp A
+\frac{8(2-\delta)  }{3 \pi} A^2 -\frac{3 }{4}  A^3  
\label{A1} 
\end{equation} 
where $\nabla^2_\perp = \partial_x^2+\partial_y^2$, $\lambda (h) =
\delta-1 - \frac{\pi^2}{4 h^2}$.  Deriving this equations we assumed
that $(2-\delta )A^2$ and $A^3$ are of the same order, i.e. $\delta
\approx 2$, however qualitatively similar equation with a different
nonlinearity can be obtained for any $\delta$ and $h$. 
Eq. (\ref{A1}) must be coupled to the  mass conservation 
equations which reads as (here we neglect contribution 
from the flux along $y$-axis  $J_y \sim \partial_y h \ll  J$): 
\begin{equation} 
\frac{\partial h}{\partial t} = 
-\frac{\partial J}{\partial x}= - \alpha 
\frac{\partial h^3 A}{\partial x},
\label{conser}
\end{equation} 
where $J$ was calculated from Eq. (\ref{j}) and $\alpha= 
2(\pi^2-8)g\sin\varphi /\eta \pi^3$.
Taking into account that variations in $h$ also change local surface
slope, we adopt $\delta= \delta_0 - \beta h_x$ with 
$\beta=1/(\phi_1-\phi_0)$. 

We studied Eqs. (\ref{A1},\ref{conser}) numerically. The simulations
were performed in fairly large systems, $400$ dimensionless units in
$x$-direction (downhill), and $200$ units in $y$-direction, with the
number of grid points $1200\times 600$ correspondingly.  
As initial conditions  we used uniform static layer:
$h=h_0, A=0$.  We triggered avalanches by a localized perturbation
introduced near the point $(y,z)=(L_y/4,L_z/2)$.  Close to the solid line
in Fig. (\ref{stab}) we indeed observed avalanches propagating only
downhill, with the shape very similar to the experimental one.  The
avalanche leaves triangular trace with the opening angle $\psi$ in which
the layer thickness $h$ is decreased with respect to original value
$h_0$. At the front of the avalanche the layer depth is increased with
respect to $h$, as in experiment.  The opening angle as a function of
$\delta$ is shown in inset of Fig.\ref{stab}.

\begin{figure}[h]
\centerline{ \psfig{figure=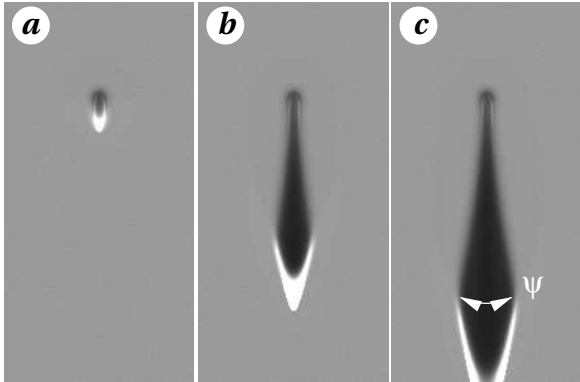,height=2.in}}
\caption{
Grey-coded images demonstrating evolution
of triangular avalanche for   $t=50 $ (a),
$t=200$ (b) and $250$ (c). White shade correspond to maximum height of
the layer, and black to minimum height.
 Parameters of Eqs. (\protect
\ref{A1},\ref{conser}) are: $\alpha=0.15, \beta=0.25, \delta=1.2$ and $h_0=3$,
point $A$ in Fig. \protect \ref{stab}. 
}
\label{Fig4}
\end{figure}

\begin{figure}[h]
\centerline{ \psfig{figure=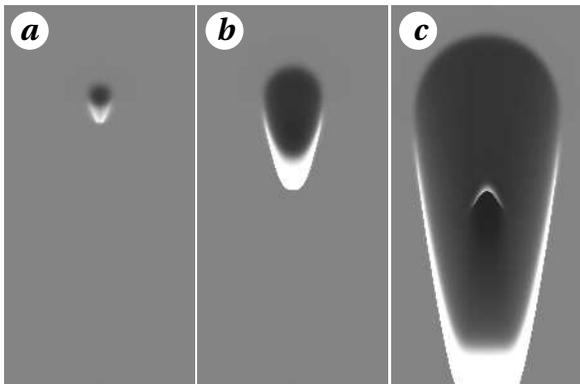,height=2.in}}
\caption{
Images of up-hill  avalanche for $t=40 $ (a),
$t=100$ (b) and $180$ (c).  Parameters of Eqs. (\protect
\ref{A1},\ref{conser}) are: $\alpha=0.05, \beta=0.25, \delta=1.07$ and
$h_0=5.5$, point $B$ in Fig.  \protect \ref{stab}.  A small secondary 
avalanche is seen on the image (c).} 
\label{Fig5} 
\end{figure}

For larger values of $\delta$ (close to dashed line in Fig.
(\ref{stab})) or for thicker layers we observed avalanches of the second
type. In this case the avalanche propagates also uphill, and contrary
to the previous case, the while avalanche zone is in motion, as new
rolling particles are constantly arrive from the upper boundary of the
avalanche zone. Sometime we observed small secondary  avalanches in the
wake of large primary avalanche, see Fig. \ref{Fig5}c. 

In conclusion, we developed a  continuum description of partially fluidized 
granular flows. Our order-parameter model captures important aspects of the 
phenomenology of chute flows  observed in recent experiments 
\cite{daerr,daerr1,pouliquen}, including
the structure of the stability diagram, triangular shape of downhill
avalanches at small inclination angles and balloon shape of uphill
avalanches for larger angles. We believe that our model can be
applicable to other granular flows including sandpiles and rotating
drums.

We thank Dan Howell, Deniz Ertas, Joe Goddard, Bob Behringer and Adrian Daerr 
for useful discussions. This research is supported by the Office of the
Basic Energy Sciences at the US Department of Energy, grants W-31-109-ENG-38, 
DE-FG03-95ER14516, and DE-FG03-96ER14592.

\end{document}